\documentclass[conference,10pt]{IEEEtran}
\IEEEoverridecommandlockouts
\usepackage{cite}
\usepackage{amsmath,amssymb,amsfonts}
\usepackage{algorithmic}
\usepackage{graphicx}
\usepackage{textcomp}
\usepackage{multirow}
\usepackage{xcolor}
\usepackage{url}
\def\BibTeX{{\rm B\kern-.05em{\sc i\kern-.025em b}\kern-.08em
    T\kern-.1667em\lower.7ex\hbox{E}\kern-.125emX}}

\usepackage{algorithm,algorithmic}

\usepackage{physics,bbding}
\usepackage{color,hyperref}
\hypersetup{colorlinks,
            breaklinks,
            linkcolor=black,
            urlcolor=black,
            anchorcolor=black,
            citecolor=black}

\begin{document}

\title{Quantum Data Center: Perspectives
}

\author{
\IEEEauthorblockN{Junyu Liu}
\IEEEauthorblockA{
\textit{The University of Chicago}\\
junyuliu@uchicago.edu}
\and 
\IEEEauthorblockN{Liang Jiang}
\IEEEauthorblockA{
\textit{The University of Chicago}\\
liang.jiang@uchicago.edu}
}

\maketitle

\begin{abstract}
A quantum version of data centers might be significant in the quantum era. In this paper, we introduce Quantum Data Center (QDC) \cite{Liu:2022ubu}, a quantum version of existing classical data centers, with a specific emphasis on combining Quantum Random Access Memory (QRAM) and quantum networks. We argue that QDC will provide significant benefits to customers in terms of efficiency, security, and precision, and will be helpful for quantum computing, communication, and sensing. We investigate potential scientific and business opportunities along this novel research direction through hardware realization and possible specific applications. We show the possible impacts of QDCs in business and science, especially the machine learning and big data industries.
\end{abstract}

\section{Introduction}
In recent years we have witnessed the vigorous development of quantum technology. Nowadays, technologies in the Noisy Intermediate Scale Quantum (NISQ) era \cite{preskill2018quantum}, where people could build intermediate-scale quantum devices and perform sophisticated experiments with a significant amount of data (see, for instance, \cite{arute2019quantum}). In the long-term future, we expect that large-scale, universal, and fault-tolerant quantum devices will emerge. The combination of quantum technologies with existing classical data science and machine learning might allow us to solve more challenging problems across science and industry. 

Data centers \cite{barroso2009datacenter} are collections of dedicated hardware dealing with large-scale data. Besides their long history from large computer rooms of the 1940s (typified by ENIAC), data centers have experienced a renaissance in the Internet age, together with the rise of cloud computing \cite{greenberg2008cost}. Thus, we expect that a quantum version of data centers should be naturally developed in order to meet the possible large-scale data processing needs of the upcoming quantum era. Our quantum version of data centers needs to have broad applications in information science in such a quantum age, including quantum computing \cite{feynman2018simulating,shor1999polynomial,grover1997quantum,preskill2018quantum}, quantum communication \cite{gisin2002quantum,cleve1999share,hillery1999quantum,kimble2008quantum,caleffi2018quantum,muralidharan2014ultrafast,muralidharan2016optimal}, and quantum sensing \cite{degen2017quantum,giovannetti2006quantum,giovannetti2011advances}.

\begin{figure}[h!]
\begin{center}  
\includegraphics[width=0.5 \textwidth]{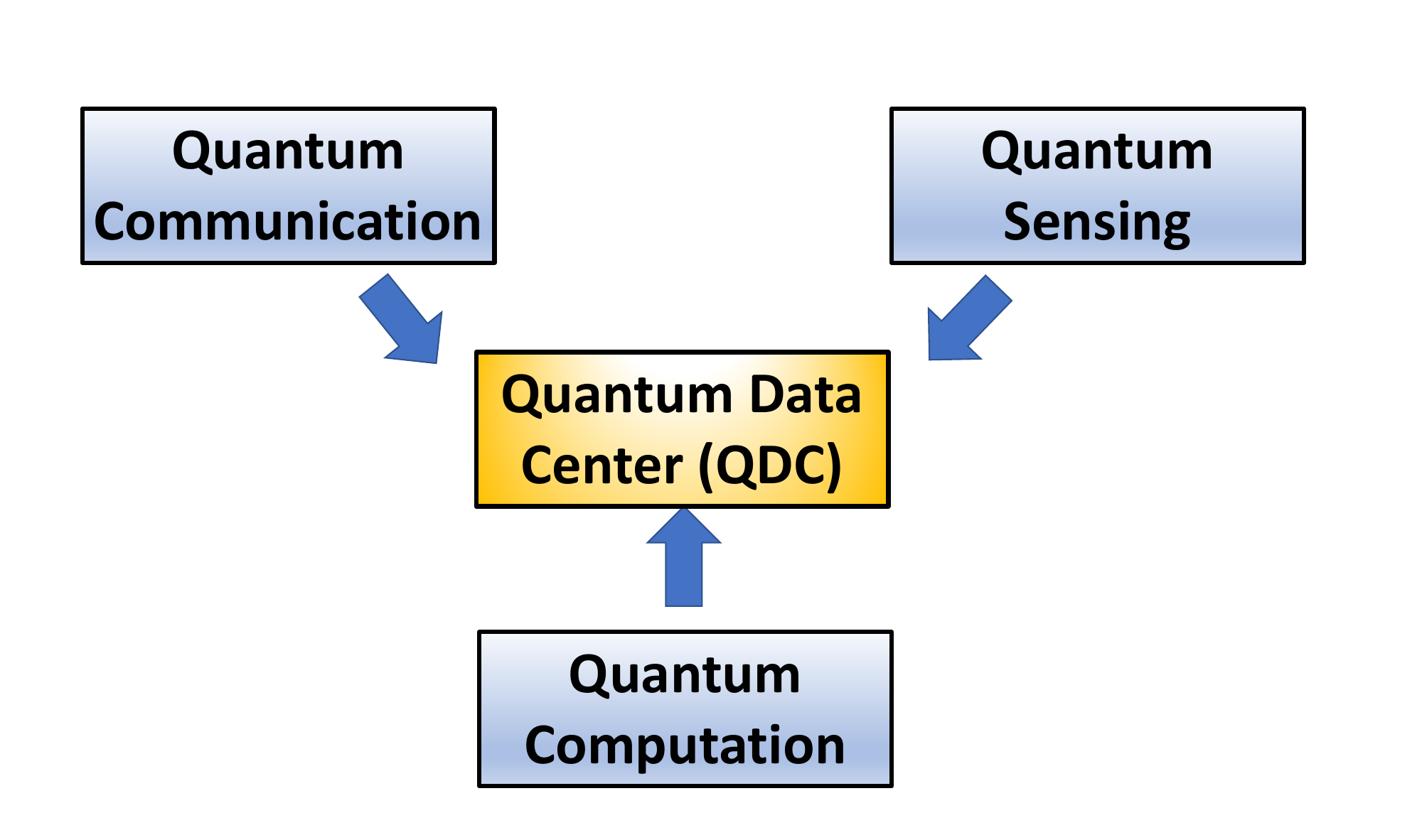}
\caption{Quantum data center (QDC) could potentially provide services about generation, processing, and application of quantum data, which could have wide applications in quantum computation, quantum communication, and quantum sensing. }
\label{fig:qdc_mean}
\end{center}  
\end{figure}

What kind of hardware form should such a quantum architecture have? Here, we propose the concept of \emph{Quantum Data Center} (QDC) \cite{Liu:2022ubu}. We point out that any QDC should include two natural parts: Quantum Random Access Memory (QRAM) \cite{giovannetti2008quantum,giovannetti2008architectures,hong2012robust,arunachalam2015robustness,hann2019hardware,di2020fault,paler2020parallelizing,hann2021resilience,connorthesis} and quantum network \cite{yin2017satellite,liao2017satellite,chen2021integrated,briegel1998quantum,duan2001long,kimble2008quantum,munro2015inside,muralidharan2014ultrafast,muralidharan2016optimal}. QRAM is a specific type of quantum memories that allows superpositions of quantum addresses and outputs, while quantum network facilitates the transmission of information in the form of qubits, between quantum processors across physical distances. We argue that a combination of QRAM and quantum networks, QDCs, could unlock the potential from quantum information science for various data-science-related tasks. Several examples of QDCs applied in quantum computing, quantum communication, and quantum sensing indicate that QDCs could provide efficient, secure, and fast services for customers (see Figure \ref{fig:qdc_mean}). 

The structure of our paper is given by the following Table \ref{table1}.

\begin{table}[ht]
\centering
\begin{tabular}{p{5.4cm}  p{ 2.5cm}}
\hline
QDC Materials & Relevant Section \\
\hline
Hardware Introduction & Section \ref{hard}\\ 
Hardware Introduction: QRAM &  Section \ref{hardqram} \\ 
Hardware Introduction: quantum networks &  Section \ref{hardnetwork} \\ 
Applications &  Section \ref{appli}\\
Applications: Quantum Computing&  Section \ref{appli:computing}\\
Applications: Quantum Communication &   Section \ref{appli:comm}\\
Applications: Quantum Sensing & Section \ref{appli:sensing} \\
Further Features of QDCs & Section \ref{further} \\
Potential Impacts & Section \ref{impact} \\
\hline
\end{tabular}
\caption{Guidance of relevant sections in the paper.}
\label{table1}
\end{table}

\section{Hardware}\label{hard}
\subsection{Quantum Random Access Memory (QRAM)}\label{hardqram}
There are many quantum algorithms running in quantum devices that could claim a significant advantage over their classical counterparts (for instance, \cite{harrow2009quantum}). However, their advantages are sometimes relying on the query models related to quantum or classical inputs. Quantum Random Access Memory (QRAM) is a physical realization of such oracles \cite{giovannetti2008quantum,nielsen2002quantum,connorthesis}. Roughly speaking, the function of QRAM is based on implementing addresses in the superposition states to ask for different elements in the database, which could be either classical or quantum. 

In the classical case, the user of QRAM will provide addresses that allow superpositions, and the QRAM database will reply the user by the entanglement states between the data and the addresses. More precisely, QRAM (with classical data) is an architecture that realize the following unitary operation
\begin{align}
\sum\limits_{i = 0}^{N - 1} {{\alpha _i}} {\left| i \right\rangle ^{{Q_1}}}{\left| 0 \right\rangle ^{{Q_2}}} \to \sum\limits_{i = 0}^{N - 1} {{\alpha _i}} {\left| i \right\rangle ^{{Q_1}}}{\left| {{x_i}} \right\rangle ^{{Q_2}}}~,
\end{align}
with the input and output qubit registers $Q_1$ and $Q_2$, the classical data $x_i$, the superposition coefficients $\alpha_i$, and the size of the data $N$. 

In the quantum case, the difference comparing to the classical ones is that now QRAM is transforming the quantum states from the memory, given the quantum addresses as inputs. More precisely, it is the following unitary operation,
\begin{align}
&\sum\limits_{i = 0}^{N - 1} {{\alpha _i}} {\left| i \right\rangle ^{{Q_1}}}{\left| 0 \right\rangle ^{{Q_2}}}\left[ { \otimes _{j = 1}^N{{\left| {{\psi _j}} \right\rangle }^{{D_j}}}} \right] \to\nonumber\\
&\sum\limits_{i = 0}^{N - 1} {{\alpha _i}} {\left| i \right\rangle ^{{Q_1}}}{\left| {{\psi _i}} \right\rangle ^{{Q_2}}}\left[ { \otimes _{j = 1}^N{{\left| {\bar \psi _j^{(i)}} \right\rangle }^{{D_j}}}} \right]~,
\end{align}
with the input and output qubit registers $Q_1$ and $Q_2$, the memory $D_j$, the quantum data $\psi_i$, the superposition coefficients $\alpha_i$, the size of the data $N$, and the notation
\begin{align}
\left| {\bar \psi _j^{(i)}} \right\rangle  = \left\{ \begin{array}{l}
\left| 0 \right\rangle {\text{ for }}i = j\\
\left| {{\psi _j}} \right\rangle {\text{ for }}i \ne j
\end{array} \right.~.
\end{align}
QRAM is designed to have efficient time cost: the unitary of QRAM could be constructed using $\mathcal{O}(\log N)$ circuit depth with $N$ qubits, under the assumption where the large-scale quantum processors scale as $\mathcal{O}(\text{polylog }N)$ \cite{giovannetti2008quantum}. Currently, there are several proposals that provide us guidances towards practical realizations of QRAM. With the help of quantum routers, one could design the fanout QRAM and the bucket-brigade QRAM architectures as two possible realizations of the QRAM unitary \cite{giovannetti2008quantum,nielsen2002quantum,giovannetti2008architectures,arunachalam2015robustness,hong2012robust,arunachalam2015robustness}. The QRAM architecture could be potentially realized in the cavity quantum electrodynamics (cavity-QED) \cite{giovannetti2008architectures,moiseev2016time}, circuit quantum electrodynamics (circuit-QED) \cite{kyaw2015scalable}, hybrid quantum acoustic systems (circuit quantum acoustodynamic systems, cQAD) \cite{hann2019hardware}, Rydberg atoms \cite{hong2012robust}, and the photonic integrated circuit (PIC) architecture \cite{chen2021scalable}, although a full experimental realization of QRAM does not exist at the moment. Moreover, a related construction of quantum memory, which is so-called Quantum Read-Only Memory (QROM), has been proposed and discussed with near-term applications, where a hybrid form of QRAM and QROM might be useful in order to balance the overheads between hardware and time \cite{babbush2018encoding,low2018trading,berry2019qubitization,di2020fault,di2019methods,hann2021resilience,connorthesis}.

\subsection{Quantum Networks}\label{hardnetwork}
Quantum networks could serve as important elements of quantum computing, quantum communication, and quantum sensing tasks. For more than two decades, tireless efforts around the world have led to significant progress in entangling quantum nodes and ultimately building global quantum networks. Especially, amazing theoretical and experimental works have been performed on quantum satellites \cite{yin2017satellite,liao2017satellite,chen2021integrated} and quantum repeaters \cite{briegel1998quantum,duan2001long,kimble2008quantum,munro2015inside,muralidharan2014ultrafast,muralidharan2016optimal}, etc,. More generally, people propose the idea of quantum internet \cite{kimble2008quantum}, where one could potentially accomplish tasks that are not possible in classical physics \cite{gottesman1999demonstrating,preskill1999plug}. We recommend \cite{wehner2018quantum} towards an overview of research about quantum networks. 

\begin{figure}[h!]
\begin{center}  
\includegraphics[width=0.5\textwidth]{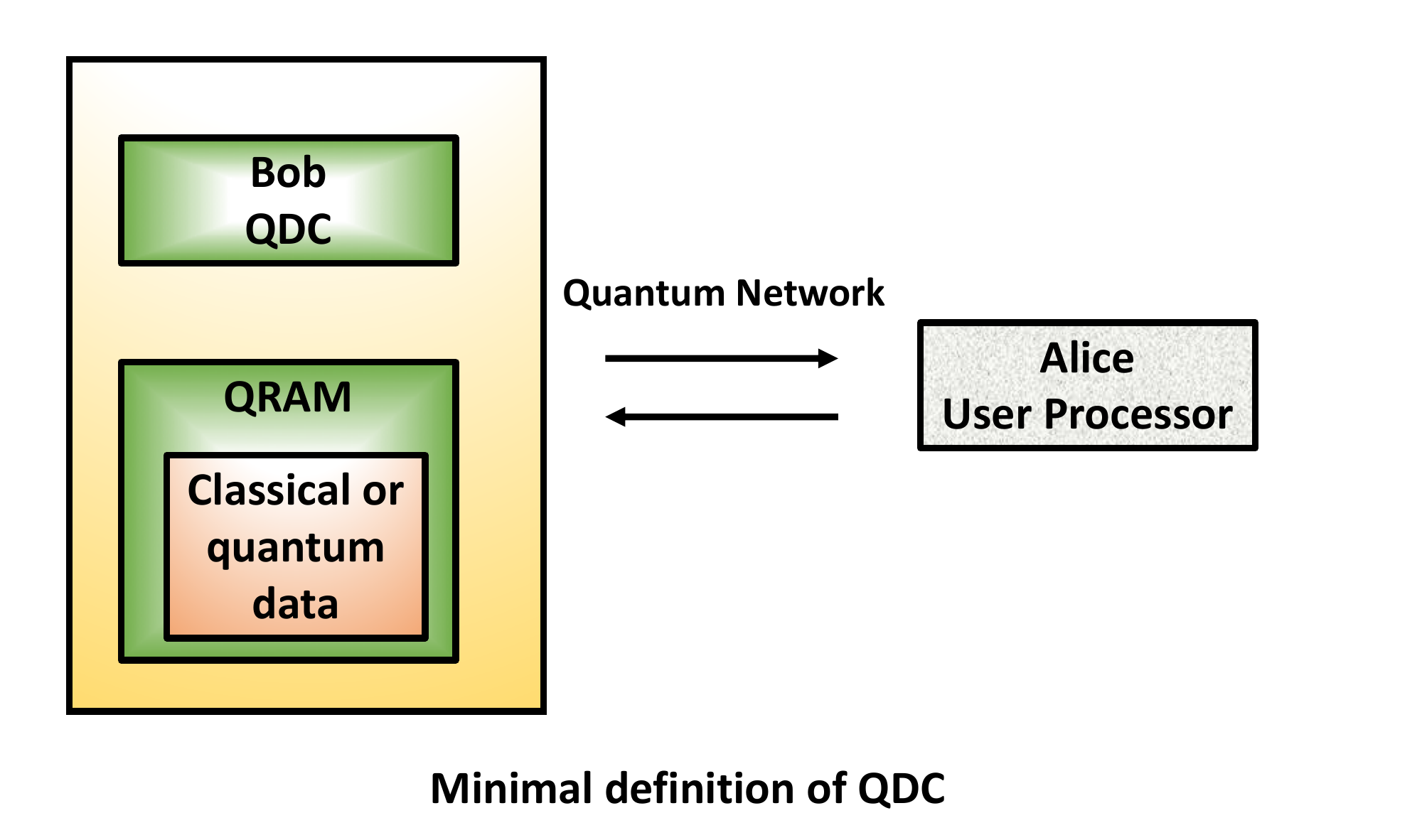}
\caption{The minimal definition of QDC contains the quantum network and QRAM. The data stored in QRAM can be either classical or quantum.}
\label{fig:qdc_minimal}
\end{center}  
\end{figure}

Since quantum networks could teleport quantum states directly, we argue that QRAM could be distributed combining quantum networks \cite{jiang2007distributed,monroe2014large}, which could form the minimal definition of QDCs. Besides QRAM which will allow efficient implementations of oracles and provide useful interfaces between classical and quantum processors, security could be enhanced inherited from function of quantum networks. See Figure \ref{fig:qdc_minimal} for an illustration.  

\section{Applications}\label{appli}
To make our claim more precise, we discuss applications of QDCs in different subjects of quantum information science, where QDCs could take specific forms according to the minimal definition. In total, we provide six applications, and a guidance is provided in Table \ref{tablecomputing}. Note that, one might be curious about how QDCs are different from a generalized version of quantum computers. Here, we should clarify that, of course, we could develop a Universal Quantum Computing (UQC) device that is associated with QDC. However, it is not necessary, and our QDC construction could directly serve remote users with their own quantum computation architectures. In fact, some of our examples mentioned later do not require UQC power for QDC. 

\begin{table*}
 \centering 
\label{tablecomputing}
\begin{tabular}{p{5.4cm}  p{ 5.4 cm} p{ 2.5cm} }
\hline
QDC Application & Subject & Relevant Section \\
\hline
Offloading magic state distillation& Quantum Computing & Section \ref{tgate} \\ 
Oracle implementation & Quantum Computing &  Section \ref{oracle}\\ 
Quantum private queries & Quantum Communication & Section \ref{qpq} \\ 
Blind quantum computing$^*$& Quantum Communication & Section \ref{blindQC}\\
Multi-party private communication & Quantum Communication &  Section \ref{multi-party}\\
Distributed sensing & Quantum Sensing & Section \ref{distributed_sensing} \\
\hline
\end{tabular}
\caption{Example applications of QDCs and their corresponding sections in this manuscript. Note that among these applications, only one---Blind quantum computing labeled with $*$---necessarily requires that the QDC be capable of universal quantum computing (UQC).}
\end{table*}

\subsection{Quantum Computing}\label{appli:computing}
\subsubsection{QDCs as $T$-gate resources}\label{tgate}
Quantum error correction is considered to be a crucial part of universal quantum computing. However, the leading paradigm for quantum error correction, the two-dimensional topological codes (surface codes, for instance), cannot directly execute non-Clifford gates, like $T$-gates through logical Clifford operations. To prepare efficient magic states as resource states of $T$-gates with high fidelity, people usually use the magic state distillation protocol \cite{bravyi2005universal,bravyi2012magic}. Thus, $T$-gates are expensive in fault-tolerant quantum computing, and reducing $T$-gate counts might be important in the design of quantum algorithms \cite{babbush2018encoding}. 

Thus, it is natural to think about centralizing the production of $T$-gates or magic states through a QDC, which could provide the first example in this paper as applications of QDCs in quantum computing. In this setup, we consider a general quantum algorithm for the user who wishes to offload large amounts of magic state distillation, to the QDC. Since we use query-based quantum algorithms, we make use of QRAM. Since we need to teleport quantum states between the user and the QDC, we need quantum networks. 

Such construction of QDCs might lead to significant advantages compared to the case without QDCs. With data size $N$, one could argue that the user has to distill $\mathcal{O}(\sqrt{N})$~\cite{low2018trading} magic states or more. However, if we have a QDC, which is a large resource of magic states, the user does not need to count for the cost of the query to the QDC. In this case, the communication cost is needed, which will only scale as $\mathcal{O}(\log N)$, which is counted from implementing the QRAM unitary. Moreover, the user does not need to introduce ancillary qubits that could potentially happen in the query-based algorithm \cite{low2018trading,berry2019qubitization,di2020fault} since we outsource the ancillary hardware to the QDC. In fact, when we outsource the full query, this approach could be exponentially more efficient than the naive approach if we measure the communication cost (see \cite{Liu:2022ubu} for more technical details). 

\begin{figure}[h!]
\begin{center}  
\includegraphics[width=0.5\textwidth]{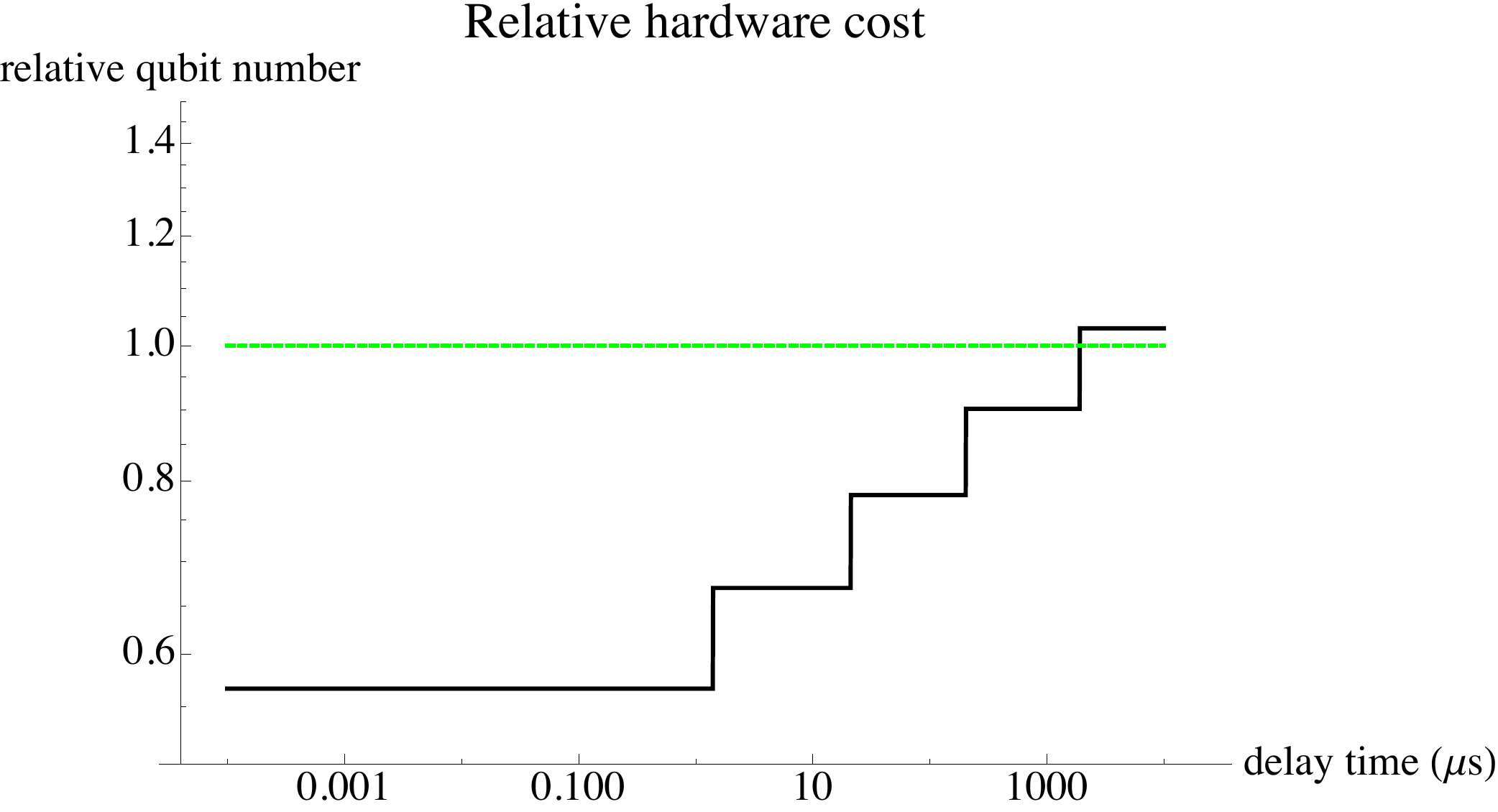}
\caption{QDC-assisted relative qubit resource costs depending on the delay time. Relative qubit number means $\frac{\text{qubit number with QDC}}{\text{qubit number without QDC}}$, given by the black line. The green dashed line gives the critical value where the relative qubit number is 1. More details are given in our technical paper \cite{Liu:2022ubu}.}
\label{fig:magiccost}
\end{center}  
\end{figure}

Figure \ref{fig:magiccost} provides an example for quantitative estimations of resources saved by QDCs in the protocol. Here we consider an example within the framework of \cite{litinski2019game} about magic state distillation. We consider a general quantum algorithm with 100 qubits and we need $10^8$ $T$-gates when we represent the circuits as Clifford and $T$-gates. Moreover, we assume the physical error rates of $p=10^{-3}$ and that the failure probability of the whole algorithm is $<1\%$, with more technical details given in \cite{Liu:2022ubu}. We find that one could define the \emph{delay factor} and the associated \emph{delay time} which quantifies the delay due to the QDC protocol including quantum communication. The discontinuous jumps in Figure \ref{fig:magiccost} are obtained from discrete changes of surface code protocols. We find that in this example, if we have fast communication with less delay time provided by QDC, we have a significant advantage on the relative number of qubits used by the user. This simple example also indicates a quantitative framework on optimizing the efficiency and parameter settings of QDCs \cite{Liu:2022ubu}. 

\subsubsection{QDCs as oracle resources}\label{oracle}
QDCs could serve as resources of other quantum oracles which are not $T$-gates or magic states. If the algorithm requires oracles where QRAM could potentially provide, one might consider using QDCs to perform the computation more efficiently. For instance, if one requires the following oracle 
\begin{align}
\left| G \right\rangle  = \sum\limits_{i=1}^L {{g_i}\left| i \right\rangle } ~,
\end{align}
where $g_i$s are specified with the computational basis $\ket{i}$ with $L$ terms in total, one could use QRAMs to prepare $\ket{G}$ states for quantum algorithms of the users. A typical example of this type is Hamiltonian simulation, where $\ket{G}$ provides the linear combination of unitaries (LCU) decomposition of the input Hamiltonian. With the simplest design of QDCs by QRAM and quantum networks, one could estimate the relative hardware cost comparing the case with QDCs as $\mathcal{O}(\log {L})/\mathcal{O}(L)$, which will lead to significant advantage especially if complicated Hamiltonians with non-local terms are included \cite{Liu:2022ubu}.

\subsection{Quantum Communication}\label{appli:comm}
\subsubsection{Quantum Private Query (QPQ)}\label{qpq}
In quantum communication, Quantum Private Query (QPQ) \cite{giovannetti2008queries} is a perfect existing example where QDCs could play a role. Imagine that Alice wishes to obtain some classical data that is saved in the database of Bob. Alice wishes to get the data, but she does not want Bob to know which data she is checking. Moreover, Bob wishes to protect the database, and he does not wish to send the data which is not asked by Alice. 

QPQ could work efficiently in such a case, with both privacies of Alice and Bob guaranteed. The data has to be saved in QRAM, where Alice will provide the address through quantum networks. Bob will return the output from the database, together with the address, to Alice. QPQ is designed such that, Alice will randomly choose either $\ket{i}$ or $(\ket{i}+\ket{0})/\sqrt{2}$ where $\ket{i}$ is the actual address Alice wants to check. Bob does not know what is the choice of Alice. Thus, Alice could check whether Bob is cheating by measuring the address. Using this way, both the privacies of Alice and Bob are protected. 

More technical details are given in \cite{giovannetti2008queries} and \cite{Liu:2022ubu}. Since the QPQ scheme will both use QRAM and quantum networks, one could naturally suggest Bob to build his database as a QDC.

\subsubsection{Blind Quantum Computing}\label{blindQC}
Another typical existing application of QDCs is the efficient blind quantum computing scheme in \cite{giovannetti2013efficient}. Blind quantum computing could be understood as an extension of QPQs towards universal quantum computing. Through blind quantum computing, the user Alice could perform a quantum algorithm $U$ without revealing to Bob what $U$ is. The idea is \cite{giovannetti2013efficient}, Bob could specify different algorithms as $U_i$ and store them in the database. Alice will tell him the service she needs by sending the address $\ket{i}$. Bob could apply the operation without measuring what $\ket{i}$ is. Alice could use the same strategy as QPQ by sending states in superposition, and Alice could detect the cheating by measuring $\ket{i}$. 

The spirit of the efficient blind quantum computing protocol is in parallel to QPQs. Since QPQs could be done using QDC, blind quantum computing could also be performed.

\subsubsection{Multi-party quantum private communication}\label{multi-party}
Aside from two existing examples of QDCs, we hereby present an original example, so-called \emph{multi-party quantum private communication}, associated with our technical paper \cite{Liu:2022ubu}. Private quantum communication indicates that one could teleport quantum information without revealing it. Moreover, our multi-party quantum private communication protocol guarantees that a third party can neither know what information was transmitted, nor which users were sending the information to others.

\begin{figure}
    \centering
    \includegraphics[width=0.5\textwidth]{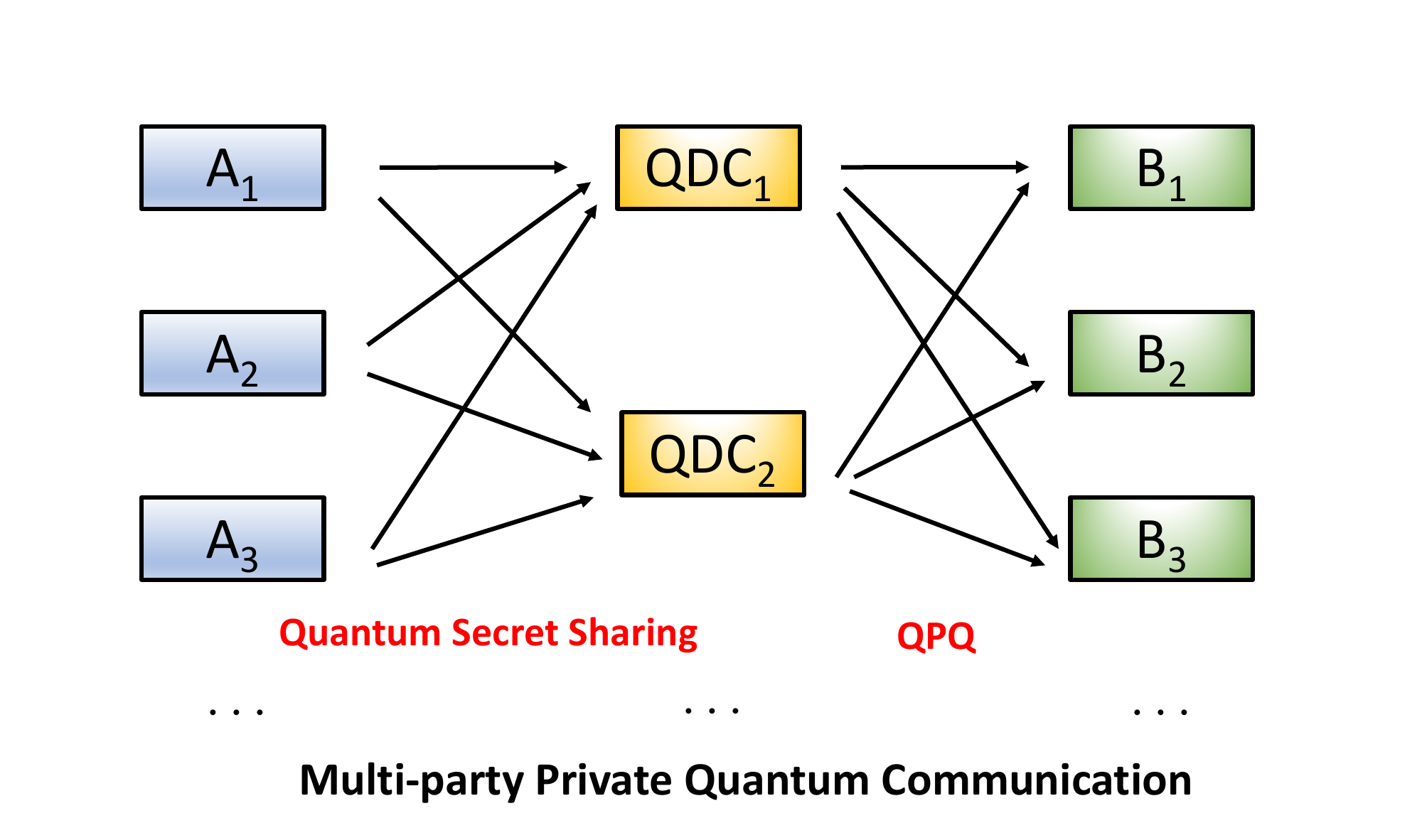}
    \caption{Multi-party private quantum communication protocol, where $A_i$s are sending information to $B_j$s without releasing the information to QDCs in the middle.}
    \label{fig:multi-party}
\end{figure}

Figure \ref{fig:multi-party} illustrates the idea of multi-party private quantum communication. Here we denote the sending users by $A_i$s, who want to send their information to $B_j$s through some untrusted QDCs in the middle. At the starting point, $A_i$s will decompose their information into several parts through a protocol called quantum secret sharing \cite{cleve1999share,hillery1999quantum}. According to \cite{cleve1999share}, one could create a $((k,n))$ threshold quantum secret sharing scheme, where each $A_i$ divide the state into $n$ parts. Any $k$ of them could recover the state of $A_i$, and any $k-1$ of them could not, as long as $n<2k$. Moreover, $A_i$s will send those divided states to many QDCs. In the end, $B_j$s will make use of QPQ to obtain the divided states in QDCs, and they can obtain the original information from $A_i$s. Using this method, one can both protect the content of the information, and the identity of senders.  

Our multi-party private quantum communication combines quantum secret sharing, QPQ and quantum teleportation. Thus, the protocol could be naturally implemented using QDCs, where we provide further detailed discussions in \cite{Liu:2022ubu}.

\subsection{Quantum Sensing}\label{appli:sensing}
\subsubsection{Data compression and distributed sensing}\label{distributed_sensing}
In our example, QDCs could be used in quantum sensing primarily through quantum data compression assisted by QRAM. Our protocol developed in \cite{Liu:2022ubu} enables QRAM and QDCs to compress the quantum data isolated in some subspaces of the whole Hilbert space. For instance, consider our quantum data is located in the single excitation space among $N$ qubits. The total Hilbert space dimension is $2^N$, and we wish to compress the full state into the $N$-dimensional subspace. One could realize the data compression task by QRAM queries. First, the QDC performs an operation to encode the location of the single excitation state into an address, where the operation could be achieved by a simple modification of QRAM designs \cite{Liu:2022ubu}. Second, we perform the QRAM query to extract the excitation state from the memory. 

The data compression scheme assisted by QRAM has direct applications in distributed quantum sensing. Physically, the $N$-qubit states could be isolated as single excitations in those tasks, and we could use quantum networks to transmit Bell pairs. If we make use of quantum data compression in QDC, we only need to transmit $\log N$ Bell pairs. A typical example is called the quantum-assisted telescope arrays \cite{khabiboulline2019optical,khabiboulline2019quantum}, where optical interferometry is used with the help of quantum networks.

\section{Further Features}\label{further}
Aside from minimal definitions, QDCs could potentially have the following features.
\begin{itemize}
\item Quantum-transduction-assisted QDC. QDCs could provide useful platforms transducing one form of quantum information to the other. Nowadays, multiple physical realizations of qubits are developing in progress, including superconducting qubits, trapped ions, neutral atoms, photonics, etc. QDCs could provide services for different quantum commercial companies with their hardware. In fact, QDCs themselves might have efficient quantum transducers between superconducting qubits and photons, when combining QRAM and quantum teleportation. See Figure \ref{fig:qdc_trans}. 
\begin{figure}[h!]
\begin{center}  
\includegraphics[width=0.5\textwidth]{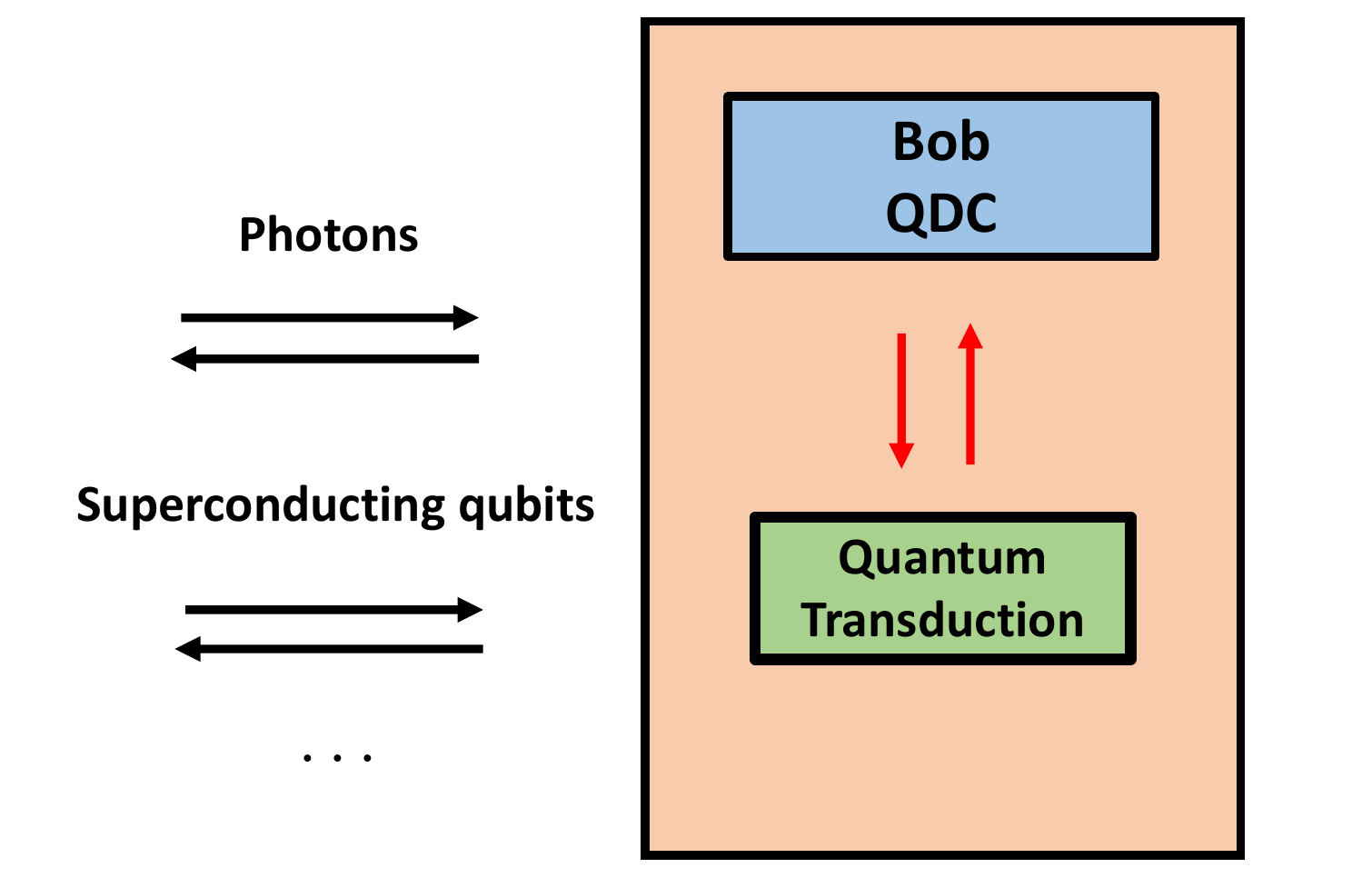}
\caption{Further directions: QDCs could be assisted by quantum transduction.}
\label{fig:qdc_trans}
\end{center}  
\end{figure}  
\item QDCs could be read-only, or could allow the users to write the corresponding data efficiently (see Figure \ref{fig:qdc_readwrite}). The definition of \emph{writing the data into QRAM} contains different situations: classical or quantum data defining different QRAM operations, while classical and quantum addressing defining how to get access to the database (see Figure \ref{fig:qdc_readwrite} for an illustration).
\begin{figure}[h!]
\begin{center}  
\includegraphics[width=0.5\textwidth]{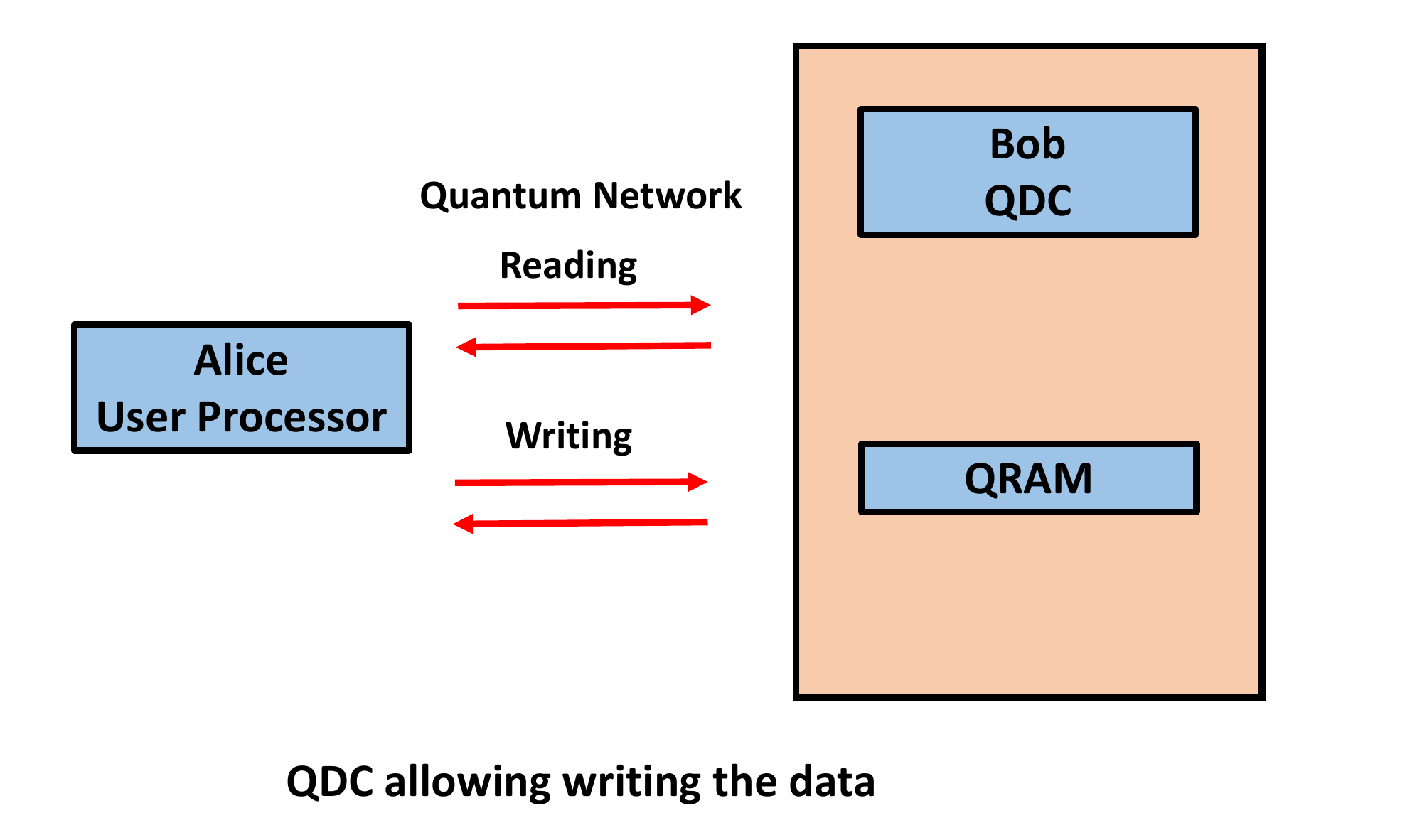}
\caption{Further directions: QDCs could not only be read-only, but also allow writing functions to change the database.}
\label{fig:qdc_readwrite}
\end{center}  
\end{figure}  
\item QDCs could be centralized or even distributed (as the modular QDC) (see Figure \ref{fig:qdc_distributed}). It could be realized by distributed quantum computation \cite{kimble2008quantum,caleffi2018quantum}. The concept of the distributed quantum computation \cite{kimble2008quantum,caleffi2018quantum} is similar to its classical analog: cluster computation. If we are able to perform quantum computation through a rapid quantum network, then we could gain more computational power assuming each node in the quantum network is a small quantum processor. In our case, we could allow QDCs to be distributive. Then, in this case (where we call it the modular QDC), we could include the function of distributed quantum computation with the help of QRAM in QDCs. Moreover, modular QDC could also be useful for communication and sensing, for instance, the quantum-assisted telescope arrays \cite{khabiboulline2019optical,khabiboulline2019quantum} and distributive quantum sensing we discuss before.  
\begin{figure}[h!]
\begin{center}  
\includegraphics[width=0.5\textwidth]{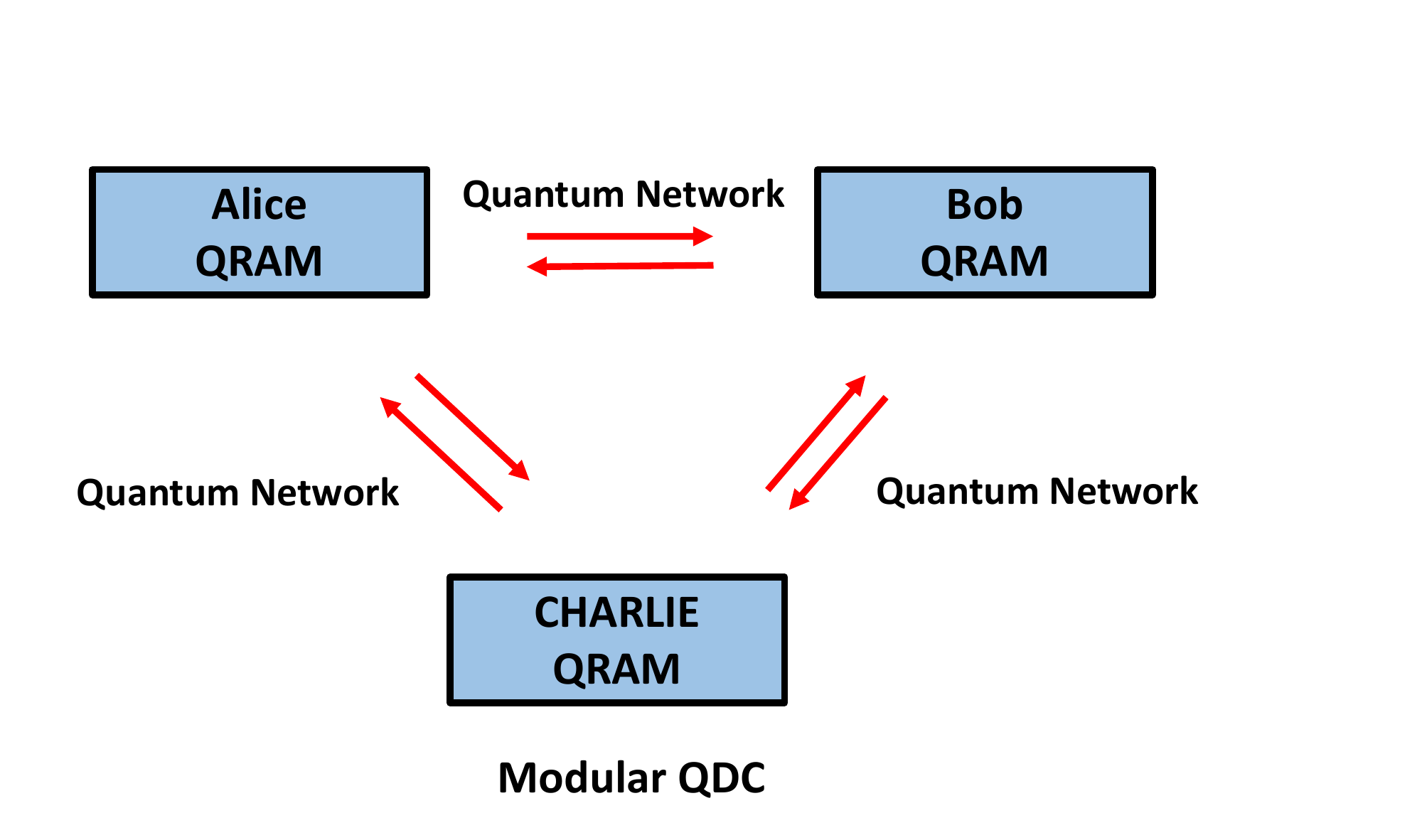}
\caption{Further directions: QDCs could be distributive and modular.}
\label{fig:qdc_distributed}
\end{center}  
\end{figure}  
\item QDCs might select suitable quantum data structures for given algorithms specified by the users. The QRAM architectures could be understood as the simplest version of the quantum data structure: the quantum version of linear lists. In classical computation, we have not only linear lists and pointers, but also linked lists, trees, graphs, stacks, queues, hash tables, etc. It will be interesting to investigate suitable quantum analogs of those data structures associated with given quantum algorithms, including insertion, deletion, search, update, traverse, and sorting \cite{fillinger2013data,Liu:2022ubu}. Those functions, if realized, could be naturally embedded into QDCs. Moreover,  it is possible that there exist quantum data structures that are from unique features of quantum mechanics, and have no analog of the classical ones. See Figure \ref{fig:qdc_qds}.
\begin{figure}[h!]
\begin{center}  
\includegraphics[width=0.5\textwidth]{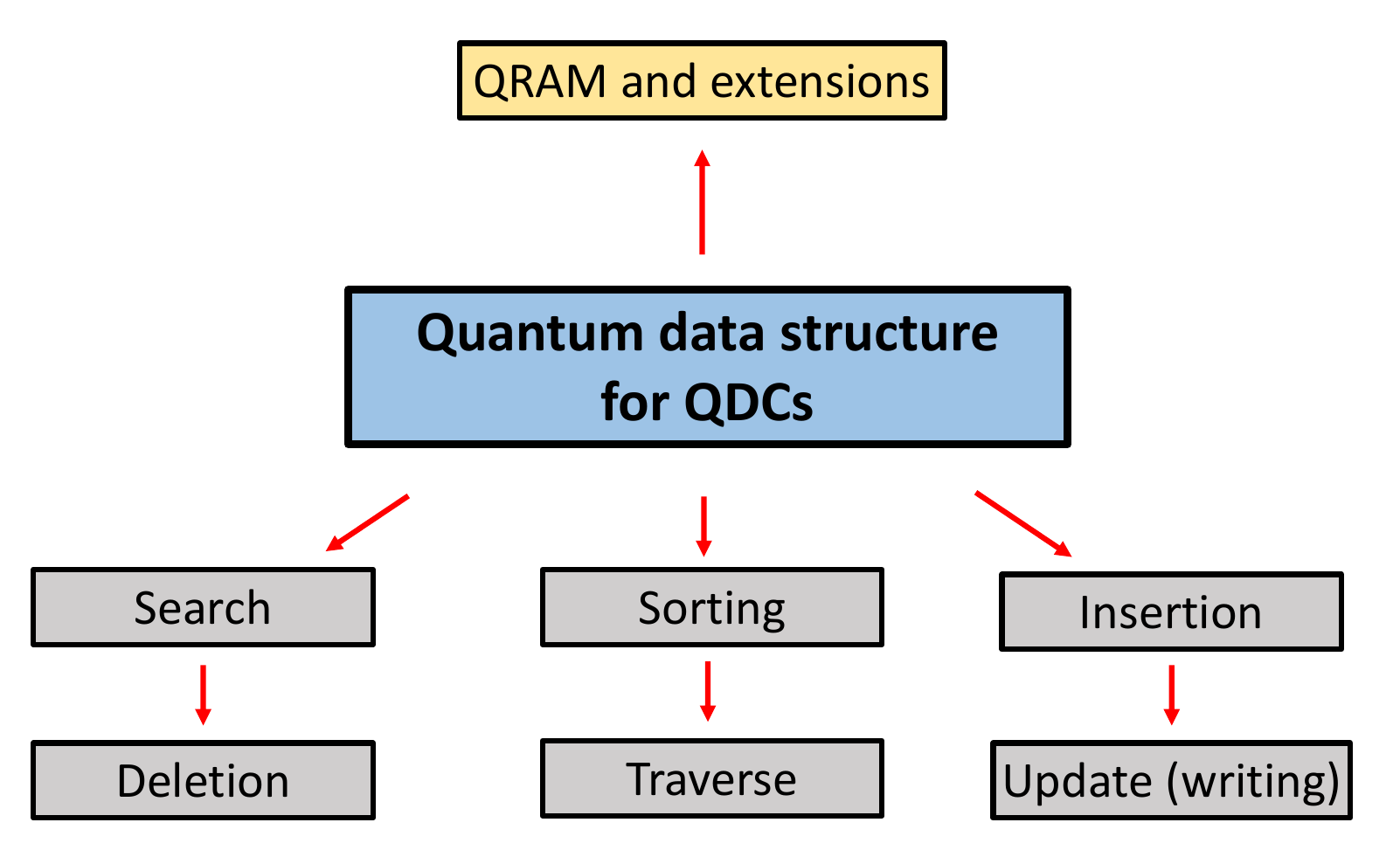}
\caption{Further directions: QDCs could provide suitable quantum data structures for given algorithms specified by the users.}
\label{fig:qdc_qds}
\end{center}  
\end{figure}  
\end{itemize}

\section{Impacts}\label{impact}
We believe that QDCs will have important applications across the industrial and scientific communities. In this section, we give a short discussion on some of their potential impacts.

\subsection{Industry} 
Although the internet allows us to work online nowadays, information and data have to be located somewhere on the earth. Thus, modern data centers are important for everyday business, including communication, finance, E-commerce, education, social networks, and many other areas. Many technology companies require their own data centers. For instance, currently (2021), Google has many data centers located in the world \cite{google}, playing an important role in realizing and maintaining their products. Some data centers are naturally merged with cloud computing and could provide cloud service for business \cite{greenberg2008cost}. 

The existence of commercial data centers has a long history, and in the quantum era, commercial QDCs could appear as a natural extension of classical data centers. We argue that QDCs could be important and beneficial for several perspectives in the industry. 
\begin{itemize}
\item Security enhancement. QDCs could provide significant security enhancement in order to protect customers' data and programs. The application of QRAM itself, associated with QPQ, could guarantee the protection of data security based on the principle of quantum mechanics. Quantum teleportation and quantum internet technologies promise to provide fast and secure information transfer against possible attacks. Thus, a combination of them, forming QDCs, could naturally satisfy and enhance the security standard of data centers. Those functions are helpful, especially for finance, E-commerce, and social networks, to maximally protect the privacy of customers. 
\item Merging with quantum industry. The development of classical data centers is together with the growth of computational power in the modern information industry, and we will expect that a similar situation will happen in the quantum era. Since the whole quantum industry is developing rapidly, quantum-assisted data centers will appear in some forms. Naturally merging with the quantum industry, QDCs could provide early-bird access to the quantum world for all technology companies interested in quantum. Moreover, QRAM associated with QDCs could provide natural access between classical and quantum industries, with quantum teleportation connecting data centers and users. 
\item Cloud services. Nowadays (2022), cloud computing has become popular and common. Several large technology companies provide different types of cloud services, for instance, Amazon \cite{amazon}, and Microsoft \cite{microsoft}. QDCs provide a natural quantum enhancement of cloud services through cloud quantum computing. Higher computational power-assisted by quantum technologies in the cloud, robust security guarantee provided by QRAM, and quantum teleportation could be delivered to users through QDCs, including public, private, hybrid, distributed, big data, High-Performance Computing (HPC) clouds for different categories of customers.    
\end{itemize}

\subsection{Science}
Science, even fundamental science, has a long history of delivering unexpected new technologies to the sociality. The World Wide Web (WWW) was born in European Organization for Nuclear Research (CERN), the world center of fundamental particle physics \cite{mcpherson2009tim}. Many aspects of modern semiconductor technologies are originated from the Apollo program, preparing and landing the first humans on the Moon \cite{hall1996journey}. Thus, the first customers of QDCs might also be from scientific collaborations as well. Here, we briefly point out several scientific collaborations that might benefit from QDCs (see discussions also in \cite{alexeev2021quantum}).
\begin{itemize}
\item Physical science. CERN and other organizations with particle collider experiments, are famously known for their large databases. For example, Large Hadron Collider (LHC) generates around $9\times 10^7 \text{GB}$ per day \cite{cern}. It is straightforward to extend their devices to QDCs from the existing databases (so-called EOS) \cite{cern}, and, in fact, it already starts \cite{cernopenlab}. Moreover, quantum sensing could be used in dark matter detection \cite{Carney:2020xol}, where QDCs could naturally provide the service by merging quantum sensing and quantum database. See also similar applications in radio astronomy \cite{Kordzanganeh:2021azm} and gravitational wave detection \cite{Korobko:2019szx}. (see also \cite{Liu:2021ohs})
\item Chemistry and materials science. Quantum computational chemistry is considered as one of the earliest near-term applications of quantum computing technologies \cite{mcardle2020quantum}. When we simulate quantum dynamics in quantum chemicals and materials from the first principle, QDC could provide a natural resource of quantum oracles. Furthermore, QDC could be a natural database storing categories of chemical data efficiently and securely. Moreover, provable quantum advantage has been recently identified \cite{huang2021quantum} in quantum machine learning algorithms for physical experiments, where QDC with QRAM might potentially provide resolutions of interfaces between quantum and classical processors. 
\item Biology, finance, and climate science. Near-term quantum algorithms might be helpful for optimization problems that could emerge in the research of biology and medicine \cite{cordier2021biology}, finance \cite{bouland2020prospects} and climate science \cite{berger2021quantum}. Optimization algorithms like Quantum Semidefinite Programming algorithms \cite{brandao2017quantum} could be implemented in the quantum devices, while oracle constructions QDC could provide might be needed to achieve provable quantum advantage. Moreover, scientific data from research in those areas could provide a list of data structures where QDCs could potentially offer \cite{cordier2021biology}.
\end{itemize}

\subsection{Machine learning and big data}
One of the most exciting applications of quantum information science might be a quantum version of machine learning or data science \cite{wittek2014quantum}. Existing data centers have significant overlaps with the machine learning community and data science, from methodology, scientific research, to commercialization. Extending classical data center technologies will have great importance to embrace fast-developing data science in a quantum scenario.    
\begin{itemize}
\item From the QRAM perspective, QRAM could be an important ingredient of quantum machine learning algorithms \cite{wittek2014quantum,biamonte2017quantum}. First, QRAM could serve as an efficient way of loading classical or quantum data in the learning problems. Making use of QRAM to obtain features from big data might be important to obtain provable quantum advantage (see \cite{kerenidis2018q,kerenidis2019quantum}). Moreover, linear algebra operations appear everywhere in machine learning and data science algorithms. The celebrated Harrow-Hassidim-Lloyd (HHL) algorithm could offer significant speedup for matrix inverses, which requires fast and efficient oracles where QRAM and QDCs could provide \cite{harrow2009quantum}. Guaranteed speedup could show up when combining QRAM and those quantum algorithms, for instance recently, \cite{zlokapa2021quantum}.  
\item From the quantum communication perspective, quantum machine learning might play an important role in the context of quantum networks. In the situation of quantum networks of sensors \cite{xia2021quantum}, quantum machine learning algorithms obtain a significant speedup for the task of classifying data (see also \cite{lloyd2021quantum,huang2021quantum}). Moreover, machine learning algorithms are proven to be useful for designing quantum networks \cite{wallnofer2020machine}. Thus, an interplay between QDCs and machine learning will be interesting and important, following the growing path of quantum communication technologies. 
\end{itemize}

\section*{Acknowledgement}
We thank Gideon Lee, John Preskill, Nicolas Sawaya, Xiaodi Wu, Xiao Yuan, Pei Zeng and Sisi Zhou for useful discussions, and Connor Hann for joining at the initial stage of this work and many valuable comments. J.L. is supported in part by International Business Machines (IBM) Quantum through the Chicago Quantum Exchange, and the Pritzker School of Molecular Engineering at the University of Chicago through AFOSR MURI (FA9550-21-1-0209). L.J. acknowledges support from the ARO(W911NF-23-1-0077), ARO MURI (W911NF-21-1-0325), AFOSR MURI (FA9550-19-1-0399, FA9550-21-1-0209), NSF (OMA-1936118, ERC-1941583, OMA-2137642), NTT Research, Packard Foundation (2020-71479), and the Marshall and Arlene Bennett Family Research Program. This material is based upon work supported by the U.S. Department of Energy, Office of Science, National Quantum Information Science Research Centers.


\bibliographystyle{ieeetr}
\bibliography{quantum.bib}

\end{document}